# Experience Deploying Containerized GenAI Services at an HPC Center

Angel M. Beltre
ambelt@sandia.gov
Sandia National Laboratories
Scalable System Software
Albuquerque, New Mexico, USA

Jeff Ogden
jbogden@sandia.gov
Sandia National Laboratories
HPC Systems
Albuquerque, New Mexico, USA

Kevin Pedretti
ktpedre@sandia.gov
Sandia National Laboratories
Scalable System Software
Albuquerque, New Mexico, USA

## Abstract

Generative Artificial Intelligence (GenAI) applications are built from specialized components—inference servers, object storage, vector and graph databases, and user interfaces—interconnected via web-based APIs. While these components are often containerized and deployed in cloud environments, such capabilities are still emerging at High-Performance Computing (HPC) centers. In this paper, we share our experience deploying GenAI workloads within an established HPC center, discussing the integration of HPC and cloud computing environments. We describe our converged computing architecture that integrates HPC and Kubernetes platforms running containerized GenAI workloads, helping with reproducibility. A case study illustrates the deployment of the Llama Large Language Model (LLM) using a containerized inference server (vLLM) across both Kubernetes and HPC platforms using multiple container runtimes. Our experience highlights practical considerations and opportunities for the HPC container community, guiding future research and tool development.

## 1 Introduction

Developers and researchers at HPC centers are increasingly seeking to deploy persistent services related to standing up local GenAI applications. Example services include Large Language Model (LLM) and Vision Language Model (VLM) inference servers, databases of various types (e.g., vector, graph, and time-series), web-based user interfaces, and agentic-AI tools. These services may be composed together and integrated with traditional modeling and simulation applications, for example to automatically generate and manage end-to-end scientific workflows, or to operate as standalone GenAI applications, such as chatbot-style virtual subject matter experts informed by site-specific data. As the breadth and variety of these services continue to evolve, enabling user self-service using technologies such as containers and container orchestration platforms becomes a key factor in improving user productivity and reducing the burden on system operators and/or administrators.

In this paper, we present the approach we are taking to enable users to self-deploy containerized GenAI services in our computing environment. Because our computing resources are limited, most notably the availability of Graphics Processing Unit (GPU) [11] resources, we have had to support both traditional HPC systems as well as Kubernetes-based container orchestration platforms, allowing users to migrate their workloads to where GPU resources are currently available. While these platforms have vastly different user interfaces for deploying containerized services, our early experience suggests a unified user interface would be feasible with appropriate tool development. We also describe how we are leveraging site-wide shared infrastructure, such as object storage services [4] and container registries, in our GenAI service deployments.

To make the discussion concrete, we focus on presenting a case study deploying the vLLM [20] inference engine serving the Llama 4 Scout [25] and Llama 3.1 405B [24] models on two HPC clusters, one with NVIDIA H100 GPUs [27] and the other with AMD MI300a [1] GPUs, and a RedHat OpenShift [33] Kubernetes-based cluster with NVIDIA H100 GPUs. While the point of the case study is not to do a thorough performance evaluation, the examples presented provide a reproducible benchmarking methodology that other sites could replicate. Additionally, the patterns presented mirror those we have encountered deploying other GenAI services. We highlight general observations and lessons learned as we step through the case study workflow.

The remainder of the paper is organized as follows: Section 2 presents the converged computing architecture we are pursuing, enabling users to self-deploy containerized GenAI services. Section 3 presents a case study deploying the vLLM inference server, providing details on each stage of the workflow. Section 4 discusses lessons learned and opportunities for future work. Section 5 presents related work and the paper concludes in section 6.

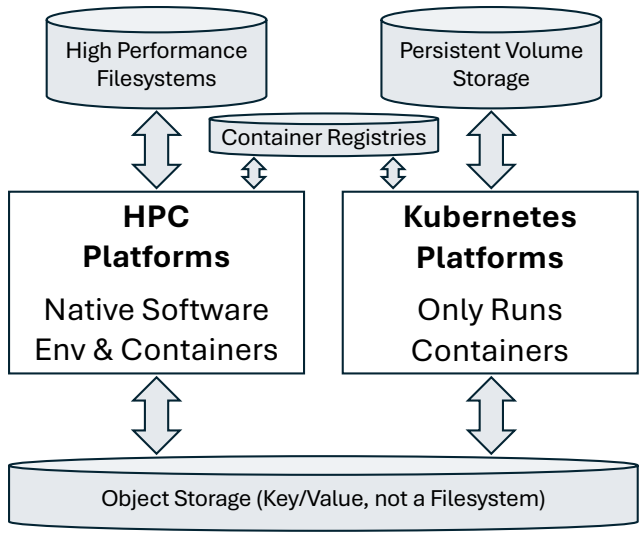

**Figure 1: Converged Computing Architecture.**

## 2 Converged Computing Architecture

Figure 1 depicts the high-level architecture of the converged computing environment we are building to support deploying GenAI services alongside traditional HPC modeling and simulation workloads. We use the term *converged* to indicate that elements of traditional HPC centers, including multiple HPC platforms and high performance parallel filesystems, are being combined with local cloud-like computing infrastructure and services, including Kubernetes container orchestration platforms, container registries, and global object storage. Each of these elements is discussed in the following subsections.

### 2.1 HPC Platforms

The HPC platforms are designed to support scientific computing applications running at large scales, for example a simulation code running in parallel on hundreds or thousands of compute nodes. The platforms include high performance networking and parallel filesystems to support this use case. In terms of the software environment, a workload manager, such as Simple Linux Utility for Resource Management (SLURM) [39] or Flux [2], is used to queue and schedule finite-duration user jobs onto the platform. Additionally, a user programming environment is provided that includes compilers, communication libraries, math libraries, debugging tools, and other software components commonly used by scientific computing applications. Lastly, HPC-specialized container runtimes are provided to enable users to build and deploy containerized jobs, for example using container images obtained from collaborators or external sources.

Sandia recently deployed two new HPC platforms—"Hops" and "El Dorado"—that include considerable GPU resources, making them an attractive place for users to run GenAI workloads such as large-scale model training. These systems were not, however, designed to support users deploying persistent services, such as GenAI inference servers, that many users are wishing to experiment with. To address this limitation, we have developed containerized workflows showing users how to easily deploy containerized GenAI inference servers and a "Compute-as-Login" mode to enable these services to be exposed to the external network, similar to network ingress in Kubernetes clusters. These capabilities are discussed in the case study presented in Section 3.

### 2.2 Kubernetes Platforms

Kubernetes [8, 19, 38] is a widely used container orchestration platform for deploying containerized applications and services. While it can support running finite-duration jobs, including batch and parallel jobs similar to those on HPC systems, it excels at deploying long-lived services and maintaining high uptime and quality of service for users. Kubernetes employs a declarative style for specifying workload deployments, where users construct deployment files that define the desired state (e.g., keep this container running, expose its service at this network ingress URL, and spawn additional instances if request latency exceeds a specified threshold). The Kubernetes control loop then works to ensure that the actual state matches the user's desired state. When containers crash or nodes go down due to system maintenance events, Kubernetes automatically re-spawns the containers on other nodes. Stateful applications are supported through the use of Kubernetes-managed persistent volumes, which users declare and associate with their containers that require storage.

Sandia has deployed several multi-tenant Kubernetes clusters using Red Hat's OpenShift platform to enable users to develop and deploy user-facing GenAI services and software-as-a-service applications. The "Goodall" Kubernetes cluster is equipped with NVIDIA H100 GPUs and InfiniBand networking [18] to support advanced GenAI use cases, while the "CEE-OpenShift" cluster provides a larger number of NVIDIA A100 and H100 GPUs for larger-scale production deployments. Additionally, there is an "Enterprise Container Platform" that offers Kubernetes for the entire lab, however, it currently lacks GPU resources.

The Kubernetes user interface for deploying workloads is quite different from that of HPC systems, making it unfamiliar to most users. To address this issue, we have created example GenAI service deployments using the Helm package manager and shared them with individual users and teams across the lab. Notably, these deployment recipes utilize the exact same containers as those used in the HPC platform deployments. The main difference lies in the syntax for specifying the deployment of the required containers.

### 2.3 Container Registries

Container registries provide a centralized location for users to archive and share container images with one another. They can also provide useful services such as access control, security scanning, and backups for the container images they store. In the converged computing architecture shown in Figure 1, container registries are used to facilitate the transfer of container images to HPC platforms and Kubernetes platforms, as well as between standalone servers and user systems where the containers are often built.

Sandia has deployed local GitLab servers, which provide per-project container registries, as well as the Red Hat Quay container registry [14]. In practice, container images usually start out as being stored in GitLab registries [17], and then once they are ready to move into production, they are additionally stored in Quay. The Quay registry has the advantage that it automaticaly performs security scanning and can be configured to automatically mirror container images between network environments.

Our experience indicates that container registries become a bottleneck when multiple nodes simultaneously pull the same container image, such as during the startup of a multi-node GenAI inference service. Optimizations such as flattening Open Container Initiative (OCI) [30] container images to single-file SquashFS [22] or Singularity Image Format (SIF) [37] images stored on a local filesystem can be useful techniques for avoiding the bottleneck, however, it is an extra step and isn't straightforward on Kubernetes platforms. To address this challenge, we have been exploring tool-based approaches for automatically managing these optimizations across platforms, which we discuss in Section 4.

## 2.4 Object Storage

Finally, a site-wide object storage [4] service provides users with a mechanism for storing key-indexed data objects that are accessible from all computing platforms. This is particularly useful for storing large data objects, such as GenAI models or mesh files, in a globally accessible location. While the HPC and Kubernetes platforms have their own local storage systems, these are generally not mounted externally due to security concerns, making object storage a valuable substrate for accessing data "blobs" across systems.

Users interact with the object storage service through a web-based RESTful API or client libraries available for all major programming languages. The ubiquity of these interface options, which do not rely on filesystem bind mounts, makes object storage particularly attractive for use within containers, especially when running on Kubernetes platforms.

Sandia has deployed approximately 30 PB of S3 object storage split across its Albuquerque and Livermore sites. Objects are globally accessible and can be automatically duplicated across sites for high availability. The Albuquerque S3 servers have 16 x 25 Gbps connection to the network, or 400 Gbps total.

The increased usage of object storage due to GenAI workloads has helped uncover several network bottlenecks between the computing platforms and S3 storage. For example, in one case, the bandwidth from Hops compute nodes to S3 storage was improved by an order of magnitude by making a simple network routing change. Going forward, this experience is informing how we architect and configure future computing platforms.

## 3 Case Study: GenAI Inference Serving

This section presents a case study demonstrating the deployment of GenAI inference services on HPC and Kubernetes platforms. We cover the complete end-to-end workflow, including downloading models from upstream sources, storing them in local object storage, deploying the inference servers, enabling off-platform access, and benchmarking performance. The workflow is fully containerized and designed to operate entirely disconnected from the external internet, with the exception of the initial model download. Each step in the workflow is described in the following sections.

### 3.1 Downloading Models

The first step of the workflow is to download models from their upstream sources. This is most commonly from Hugging Face, which has become a popular platform for vendors and researchers to release new open-weight models, such as Meta's LLaMA and OpenAI's gpt-oss models [31].

While many GenAI tools, including the vLLM inference server that we use, integrate directly with Hugging Face and will automatically download models on first use, we have found that for offline inference serving (e.g., when disconnected from the internet), it is convenient to download a model's complete Git repository. This ensures that all artifacts are obtained, including the model's LICENSE file and other documentation, which can then be stored together in local storage.

Figure 2 shows an example command for downloading a model using the Alpine Linux Git container [3]. While we could have logged into one of the HPC platforms and used the Git tool installed there, this can be problematic because of Git configuration differences across platforms (e.g., large-file support). Using a containerized approach helps with reproducibility and allows the same container image to be used on HPC and Kubernetes platforms.

```
podman run
  --volume ./cert.pem:/etc/ssl/cert.pem
  --volume ./models:/git/models
  --workdir /git/models
  alpine/git clone
    https://${USER}:${TOKEN}@huggingface.co/$MODEL
```

Figure 2: Example model download command.

After downloading a model, Figure 3 shows an example command for storing the model in local object storage using the Amazon AWS client container [36]. This step is important so that the model is available on platforms that do not mount the HPC filesystems, for example the Kubernetes platforms and individual user systems. This also ensures the models remain available when HPC filesystems are down for maintenance.

There are several nuances in the example container run command shown in Figure 3 that can be difficult for users to deal with. For example, whether the AWS_REQUEST_CHECKSUM_CALCULATION environment variable setting is required depends on the version of the AWS client container and the S3 service implementation. Additionally, the command is lengthy and must be modified depending on the container runtime being used (e.g., Podman [34], Apptainer [21], or Kubernetes), making it difficult for users to construct correctly. These differences provide motivation for developing tools to help users with managing these container deployment challenges, which we discuss in Section 4.

### 3.2 Deploying Models

The next step in the workflow is to deploy the models for others to use. For this, we use the vLLM inference server, which provides an OpenAI-compatible API endpoint that many GenAI tools and applications can integrate with. vLLM (short for "virtual" LLM) leverages optimizations inspired by operating system virtual memory management, and it has emerged as a popular tool with a highly active

```
podman run
 -e AWS_ACCESS_KEY_ID=${S3_ID}
 -e AWS_SECRET_ACCESS_KEY=${S3_SECRET}
 -e AWS_ENDPOINT_URL=${LOCAL_S3_SERVICE}
 -e AWS_REQUEST_CHECKSUM_CALCULATION=when_required
 -e AWS_MAX_ATTEMPTS=10
 --volume ./models:/aws/models
 amazon/aws-cli s3 sync
   ./models/$MODEL
   s3://huggingface.co/$MODEL
   --exclude ".git*"
```

Figure 3: Example model upload to local S3 command.

developer community. New GenAI models are generally supported by vLLM on day one or shortly after their initial release.

The vLLM project distributes official container images for each release via Docker Hub [16], which we mirror in local container registries. Figure 4 shows an example command for deploying vLLM on HPC platforms with NVIDIA GPUs using Podman. We use a similar command to deploy vLLM on HPC platforms with AMD GPUs, substituting the vLLM container image for a version built by AMD for ROCm GPUs.

There are many configuration options available for deploying vLLM. Many of the environment variable settings in Figure 4 are used to disable internet access, which is required for offline serving. Other settings depend on the model being deployed and the target platform. For example, the Llama-4-Scout model's default context window size of 10M tokens is far too large for the amount of memory available on the Hops HPC platform (4 x 80 GiB H100s per node), hence the "--max-model-len" option is needed to reduce memory requirements.

Further complications have arisen due to differences in the container runtimes available or desired to be used on different computing platforms. Figure 5 shows an example vLLM deployment using Apptainer, mirroring similar functionality as the Podman example. Aside from syntax differences, the two container runtimes have different default semantics for the execution environment they present to containers. The vLLM container assumes it is being deployed in an isolated environment running as "root" inside the container, while Apptainer, by default, runs the container as the calling user and automatically maps in their home directory. These differences cause the vLLM container to crash at startup using Apptainer's default configuration.

Lastly, deploying to the Kubernetes platforms requires a significantly different approach compared to the HPC platforms. Initially, we had developed our own custom deployment files, however, we have since migrated to using the recently added Helm chart provided by the upstream vLLM project. This chart takes care of the details of provisioning storage via a persistent volume claim, downloading the model from object storage (using the same AWS client container as Figure 3), and deploying the vLLM container. Users fill out a single YAML file with their desired configuration, for example as shown in Figure 6, and then initiate the deployment on the (remote) Kubernetes cluster using the "helm install" command.

```
podman run
 --rm
 --name=vllm
 --network=host
 --ipc=host
 --entrypoint=vllm
 --device nvidia.com/gpu=all
 -e "OMP_NUM_THREADS=1"
 -e "HF_HUB_ENABLE_HF_TRANSFER=0"
 -e "HF_HUB_DISABLE_TELEMETRY=1"
 -e "VLLM_NO_USAGE_STATS=1"
 -e "DO_NOT_TRACK=1"
 -e "HF_DATASETS_OFFLINE=1"
 -e "TRANSFORMERS_OFFLINE=1"
 -e "HF_HUB_OFFLINE=1"
 -e "VLLM_DISABLE_COMPILE_CACHE=1"
 --volume=./models:/vllm-workspace/models
 --workdir=/vllm-workspace/models
 ${LOCAL_REGISTRY}vllm/vllm-openai:v0.9.1 serve
   meta-llama/Llama-4-Scout-17B-16E-Instruct
   --tensor_parallel_size=4
   --disable-log-requests
   --max-model-len=65536
   --override-generation-config='{"
       attn_temperature_tuning": true}'
```

Figure 4: Example model deploy with Podman command.

```
apptainer exec
 --fakeroot
 --writable-tmpfs
 --cleanenv
 --no-home
 --nv
 -e "HF_HOME=/root/.cache/huggingface"
 -e "OMP_NUM_THREADS=1"
 -e "HF_HUB_ENABLE_HF_TRANSFER=0"
 -e "HF_HUB_DISABLE_TELEMETRY=1"
 -e "VLLM_NO_USAGE_STATS=1"
 -e "DO_NOT_TRACK=1"
 -e "HF_DATASETS_OFFLINE=1"
 -e "TRANSFORMERS_OFFLINE=1"
 -e "HF_HUB_OFFLINE=1"
 -e "VLLM_DISABLE_COMPILE_CACHE=1"
 --bind ./models:/vllm-workspace/models
 --cwd /vllm-workspace/models
 vllm-cuda.sif vllm serve
   meta-llama/Llama-4-Scout-17B-16E-Instruct
   --tensor_parallel_size=4
   --disable-log-requests
   --max-model-len=65536
   --override-generation-config='{"
       attn_temperature_tuning": true}'
```

Figure 5: Example model deploy with Apptainer command.

### 3.3 Accessing Models

Once the vLLM inference server startup has completed, which can take 30 minutes or more for large models, the API endpoint is ready to begin accepting requests from users. However, in order for the

```
# -- vLLM Image configuration
image:
  # -- Container image name
  repository: "vllm/vllm-openai"

  # -- Container tag / vLLM version
  tag: "v0.9.1"

  # -- Container launch command
  command: ["vllm", "serve", "/data/",
           "--host", "0.0.0.0", "--port", "8000",
           "--served-model-name", "meta-llama/
               Llama-4-Scout-17B-16E-Instruct",
           "--tensor-parallel-size=4",
           "--disable-log-requests",
           "--max-model-len=65536"
          ]

  # -- Environment variables
  env:
    - name: HOME
      value: "/data"
    - name: HF_HOME
      value: "/data"
    - name: HF_HUB_DISABLE_TELEMETRY
      value: "1"
    [...]
```

**Figure 6: Example model deploy config for Kubernetes.**

service to be accessible from outside the computing platform where it is running, some form of network ingress routing is required.

On the HPC platforms, this can be accomplished using one of two approaches depending on the desired use case. When only a single user requires access, an SSH tunnel can be created between the user's system and the compute node where the vLLM service is running. For example, a user could create an SSH tunnel by running a command like "ssh -L 8000:compute-node:8000 -N -f login-node" on their system to forward port 8000 (vLLM's default) from their system, through the HPC system's login node, to the vLLM server running on the compute node. This allows them to connect GenAI applications and tools running on their local systems to the inference services running on the HPC platform.

For use cases where multiple users will need to access the GenAI service or where it needs to run persistently (beyond the HPC platform's normal job duration limit), we have developed a *Compute-as-Login* (CaL) mode of operation. This mechanism allows compute nodes that are not physically connected to the external network to be reconfigured to operate as interactive login nodes and routed externally via system software reconfiguration. An NGINX [12] proxy running on a platform service node is used to route external traffic arriving at a specified port, through the cluster's internal network, to the compute node running the target GenAI service. This provides similar functionality to ingress routing available on Kubernetes clusters. Once a CaL resource is provisioned to a user by a system operator, the user is able to develop and re-deploy services as needed on their own.

On Kubernetes platforms, the vLLM Helm chart can be configured to enable secure ingress routing. Once established, external network traffic arriving at a given URL is routed to a cluster node that is running a vLLM instance. If vLLM containers crash (e.g., due to a memory leak bug) or are migrated between nodes for system management reasons, Kubernetes automatically takes care of restarting the container and updating the ingress routes. This is an advantage compared to CaL mode on HPC platforms, however similar functionality can be recreated by users with techniques like using cron jobs and deploying their own request routers.

Figure 7 shows an example of how to use the `curl` command to send a GenAI inference query request to a vLLM instance via the OpenAI API. Sending requests like this can be useful for debugging purposes or making one off queries.

```
curl http://localhost:8000/v1/chat/completions
  -H "Content-Type: application/json"
  -H 'Authorization: Bearer secret-api-key'
  -d '{
  "model":
    "meta-llama/Llama-4-Scout-17B-16E-Instruct",
  "messages": [{"role": "user", "content":
    "How long to get from Earth to Mars?"}],
  "temperature": 0.7
}'
```

**Figure 7: Example inference query using `curl` command.**

### 3.4 Benchmarking Models

To do more extensive performance evaluation of a deployed GenAI model inference service, vLLM includes a set of benchmarking scripts. Figure 8 shows an example of running vLLM's included `benchmark_serving.py` script configured to send a stream of user requests to a target inference service and measure performance metrics such as the response generation throughput and latencies.

While the primary focus of this paper is not on doing a thorough performance evaluation, we present benchmarking results comparing HPC and Kubernetes platforms serving variations of the Llama 4 Scout (Scout) multi-modal GenAI model using vLLM. Scout is a relatively large model consisting of approximately 200 GiB of model weights, requiring a minimum of four GPUs to serve efficiently (e.g., with 80 GiB H100's, vLLM deployments use approximately 54 GiB/GPU to store model weights and the remainder for the kvcache used to optimize queries). We additionally evaulate a 4-bit quantized version of Scout that can fit on two GPUs. Since deploying Scout with large context windows (its maximum is 10M tokens) requires many more GPUs, so we constrain "–max-model-len" such that our deployments fit on a single node.

We use the ShareGPT [5] dataset to send a stream of real-world sampled user requests to our deployed vLLM instances. The vLLM benchmarking scrips also support other datasets, such as "random" and "user-provided", however ShareGPT seemed to provide the most realistic scenario. In our evaluations, we perform multiple runs of the benchmark sweeping the maximum request concurrency ("$batch_size") from 1 to 1024 in powers of two steps. A maximum request concurrency of 1 means that a single request at a time is sent to the inference service, while a batch size of 16, for example, means that up to 16 requests at a time are sent before waiting for a

```
podman run \
  --rm
  --name=vllm-bench
  --network=host
  --ipc=host
  -e "no_proxy=${no_proxy},${TARGET_SERVER}"
  --entrypoint="/bin/bash"
  --volume "./models:/vllm-workspace/models"
  --volume "./datasets:/vllm-workspace/models/
      datasets"
  --workdir="/vllm-workspace/models"
  ${REG}rocm/vllm:rocm6.4.1_vllm_0.9.1_20250702
    -c "python3 /app/vllm/benchmarks/
        benchmark_serving.py
    --backend openai-chat
    --endpoint /v1/chat/completions
    --base-url ${BASE_URL}
    --dataset-name=sharegpt
    --dataset-path=./datasets/
        ShareGPT_V3_unfiltered_cleaned_split.json
    --model
        meta-llama/Llama-4-Scout-17B-16E-Instruct
    --max-concurrency ${batch_size}"
```

Figure 8: Example model benchmarking command.

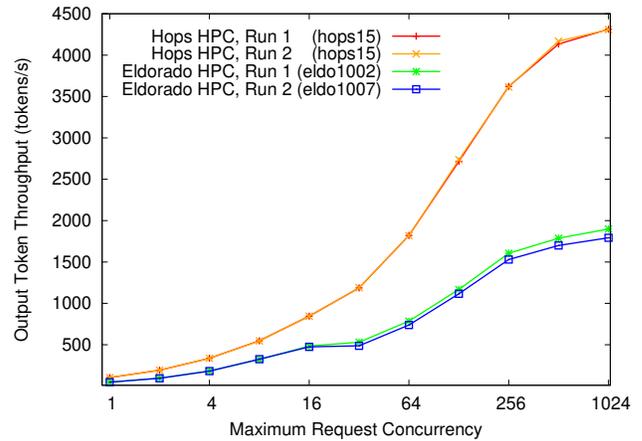

Figure 9: Hops (H100) vs. Eldorado (MI300a) performance.

response completion. This gives a good idea of the performance a single user will experience (batch 1) and the aggregate generation throughput the service can achieve for many simultaneous users (batch 1024).

*3.4.1 Hops vs. Eldorado.* Figure 9 presents results comparing the Hops HPC platform, which consists of compute nodes with 4 x 80 GiB H100 GPUs and the El Dorado HPC platform, which consists of compute nodes with 4 x 120 GiB MI300a GPUs. The key aspects of the configuration used are shown in Figure 8, with an outer loop added to sweep the batch size. The full commands are provided in the artifacts that accompany the paper. Each data point in the figure represents a single run of the benchmark consisting of 1000 user queries. For the Hops platform, with a batch size of one, the benchmark requires approximately 30 minutes to complete, while with a batch size of 1024 (an effectively unlimited max request concurrency), the same workload runs in approximately 1 minute. Each curve in the figure represents a series of benchmark runs against a unique vLLM instance running on one of the compute nodes of the respective platform, labeled in the figure key.

As can be seen in the figure, run to run variability across vLLM instances is relatively low. The Hops platform achieves a single query (batch 1) generation rate of 103 tokens/second and a maximum throughput of 4313 tokens/second (batch 1024). The El Dorado platform, in contrast, achieves a single query generation rate of 48 tokens/second and maximum throughput of 1899 tokens/second (batch 1024). While this is a significant difference between platforms, note that these are unoptimized runs using more or less default vLLM configurations. The vLLM community and vendors are achieving rapid performance gains through ongoing performance optimizations and inference engine improvments.

*3.4.2 Hops vs. Goodall.* Figure 10 presents results comparing Hops to the Goodall Kubernetes platform, which includes nodes with 2 x 94 GiB H100 NVL GPUs per node. The same benchmark configuration was used except that a smaller, quantized, version of Scout (RedHatAI/Llama-4-Scout-17B-16E-Instruct-quantized.w4a16) was used in order to fit on two GPUs, the maximum supported using a single node of Goodall.

In this case, the performance results indicate similar performance between platforms. The slight performance gain on the Goodall platform at high batch sizes can be attributed to the larger amount of HBM3 memory available (vs. 80 GiB/GPU on Hops). Additionally, the reduced maximum throughput compared to the unquantized Scout results can be attributed to only using 2 GPUs/node vs. 4 GPUs/node for the results in Figure 9.

It should be noted that the identical container image was deployed on the HPC and Kubernetes platforms. It was only the deployment mechanism that differed between platforms–a Podman command on Hops and a Helm Chart deployment for Goodall. It would be possible to abstract away these differences for users with suitable tool development, providing a common container deployment user interface, which we discuss in Section 4.

## 3.5 Multi-node Inference

Multi-node inferencing is required when a GenAI model is too large to fit within a single node, or when users wish to support larger context windows than a single node can support. In our experience, performance is generally not improved by multi-node inference, rather it is used as a way to obtain additional memory. Generally tensor parallelism is used within a node, due to higher communication requirements, and pipeline parallelism is used between nodes.

vLLM relies on Ray [26], a distributed computing framework for Python, to implement multi-node inference. Users first instantiate a Ray cluster on top of their underlying computing resources, and then start up vLLM inside the Ray cluster. vLLM uses Ray services to allocate GPUs and start remote processes running.

On the HPC platforms, we achieve this by deploying a multi-node job running one vLLM container per node, executing the Ray cluster startup command as its entry point. Once the Ray cluster is established, we exec into one of the vLLM containers (any works) and start the vLLM server. Figure 11 provides an example of this for Hops, which uses SLURM. The syntax for Flux on El Dorado is slightly different, but operates similarly.

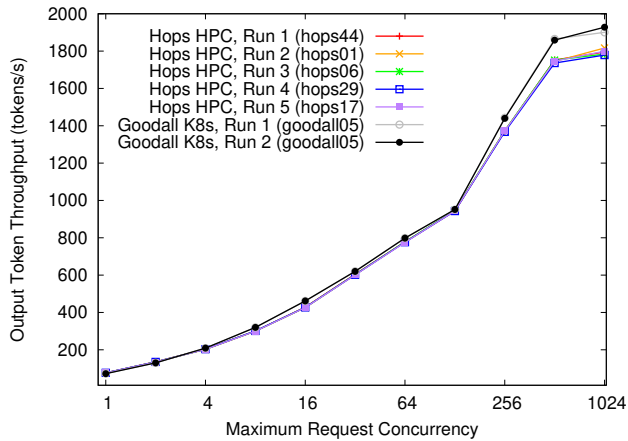

Figure 10: Hops vs. Goodall (H100-NVL) performance.

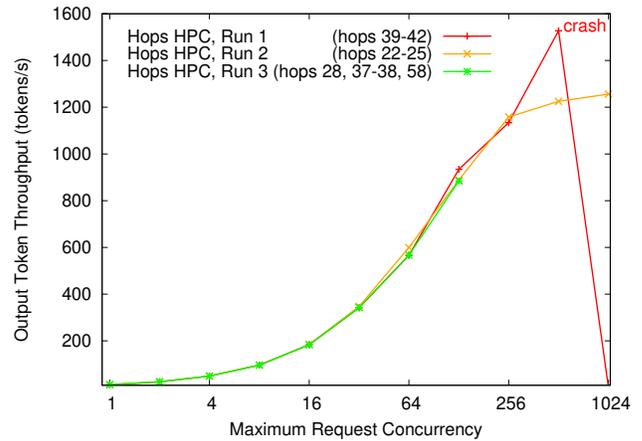

Figure 12: Hops multi-node inference performance.

```
# Start Ray Cluster
# run-cluster.sh spawns vLLM with Podman

echo "STARTING RAY HEAD on $head_node"
srun --nodes=1 --ntasks=1 -w $head_node \
    run-cluster.sh --head $head_node_ip \
    $CONTAINER_IMAGE $PODMAN_ARGS &

num_workers=$(($SLURM_JOB_NUM_NODES - 1))

echo "STARTING $num_workers RAY WORKERS"
srun -n $num_workers --nodes=$num_workers \
    --ntasks-per-node=1 --exclude $head_node
    run-cluster.sh --worker $head_node_ip \
    $CONTAINER_IMAGE $PODMAN_ARGS &

# Wait for Ray cluster to start, then spawn vLLM
```

Figure 11: Example of starting Ray cluster with SLURM.

Figure 12 presents performance results for multi-node inference with the Llama 3.1 405B model running on Hops. This model requires approximately 1 TiB of model weights, which requires 16 GPUs (4 nodes with 4 x 80 GiB H100s each). We used tensor parallelism within a node and pipeline parallelism between nodes.

Our experience has been that multi-node inference is somewhat unreliable, as seen in the figure. The first run we attempted crashed with a batch size of 512 queries. The second run completed normally and the third one was terminated early due to a scheduled system downtime. For this model, the Hops platform achieves a single query (batch 1) output generation rate of 12.5 tokens/second and a maximum throughput of 1256 tokens/second for the single successful run (run 2). Note that this run was not using InfiniBand networking, which we are still working on enabling.

## 4 Discussion

This paper has presented our experiences supporting the deployment of GenAI services within our HPC center. Qualitatively, we can say that using containers to deploy these services has been beneficial in that containers provide a straightforward mechanism for achieving reproducible software environments. AI software stacks are complex and diverse, and GenAI-related projects such as vLLM, Milvus [41], Chainlit [10], and LiteLLM [7] distribute their software in containers for reproducibility reasons, among others. Containers also allow running identical software on HPC platforms and Kubernetes, which only supports executing containers.

What we have found to be less beneficial is the complexity of the user interface for deploying containers. Our case study provides a glimpse into the complexity involved in deploying the same vLLM container with multiple container runtimes—Podman and Apptainer—and the Kubernetes container orchestration platform. Each of these systems provides a significantly different user interface for launching the exact same containerized software. This presents a high bar of entry for users, which must already deal with the complexity of the vLLM framework itself, which has many configuration options.

We see an opportunity to help users navigate the complexity of deploying containers. In particular, we believe that the following areas would benefit from increased focus by the HPC containers community:

- **Container Runtime User Interface Differences:** Each container runtime has a different user interface and default semantics. In many cases, one runtime can be configured to operate like another, as we demonstrated in our case study for Podman and Apptainer, however, this still requires significant effort by users. Container metadata could be used to encode the execution environment expectations of containerized workloads, then a tool could use this information to automatically adapt the container for different container platforms.
- **Computing Platform Differences:** Different computing platforms may require different versions of the same container, and it is difficult for users to know which to pick. For example, the upstream vLLM project only distributes CUDA containers, and users need to know where to find the ROCm optimized versions of vLLM that AMD provides. Metadata about a containerized application, for example a vLLM container "package" definition, could be used to specify which container image should be used on different computing hardware (e.g., CUDA, ROCm, or OneAPI). This is a slightly different problem than the one addressed by multi-architecture container images [32] and image labeling [9].
- **Application and Service Configuration:** Containerized applications and services can be configured to run in a multitude of ways, but there are usually only a few common high-level configurations. For example, the vLLM container can be deployed in an internet enabled way, which may require that site-specific web proxies and certificates be installed, or in an offline mode where access to the external internet is disabled, as our case study showed. Additionally, the vLLM container must be deployed differently for single-node and multi-node scenarios, which first requires that a Ray cluster be instantiated. These differences could be automatically handled for the user.
- **Computing Center Differences:** Computing centers sometimes favor a specific container runtime or have slightly different configurations for common site-wide services, such as object storage servers and container registries. This can frustrate users who successfully deploy their containerized workload at one site, only to discover they must significantly rework their deployment when moving to a new computing center. This undermines the portability benefits of using containers. We believe most of these site-specific differences can be captured in configuration profiles and automatically resolved by a suitable container deployment tool.

We have begun to prototype tooling to address these challenges, starting with the differences between container runtimes. One way to think of such a tool is as a package manager for deploying containerized applications and services, similar in concept to how the Spack [13] tool serves as a package manager for building and installing scientific software. It may be possible to merge these two concepts—package management for software building and container deployment—into a single tool, but first, the concepts need to be further explored and demonstrated.

## 5 Related Work

Early research [6, 40] explored HPC applications across supercomputer and cloud environemnts using containers and Kubernetes orchestration. These studies established the potential for converged computing environments, which our work extends by focusing specifically on GenAI service deployment across heterogeneous platforms and addressing practical deployment challenges in production HPC centers.

Building on these foundational convergence approaches, recent work has focused on optimizing specific aspects of containerized GenAI Inference deployment. Sada et al. [35] conducted a comparative study of ultra and high-performance accelerators across multiple vendors, including NVIDIA, AMD, and Qualcomm, deployed on a multi-tenant Kubernetes cluster. Their evaluation encountered limitations when benchmarking AMD MI300a GPUs with various Llama family models.

In this work, we extend the analysis of the related work by demonstrating deployment strategies for larger models across both traditional HPC environments and Kubernetes-based clusters. Our contribution includes a performance comparison of the platforms and example container deployment configurations adapted for different environments.

More advanced work has been focused on software frameworks and deployment architectures for inference optimization. NVIDIA's Inference Microservices (NIM) [28] introduce a profile-based architecture that enables runtime selection of optimized inference configurations. While effective for cloud deployments, NIM presents challenges in air-gapped environments where profile compatibility with specific accelerators cannot be guaranteed, and profile downloads may fail to achieve expected performance outcomes. NIM defaults to TensorRT-LLM [29] inference engines, but also includes vLLM profiles for some services. In our work, we default to upstream vLLM, as it has become ubiquitous across a diverse set of computing environments paired with specific accelerators in the industry.

Matias [23] explored how model deployment hyperparameter configurations affect inference performance and highlighted the significance of hyperparameter optimization for maximizing throughput. Matias optimized using hyperparameters such as GPUs and batch sizes for both inference engines vLLM and Hugging Face pipelines [15]. The best configurations from these optimization studies could be encoded in application-specific container deployment recipes, like we describe in our discussion. Our work additionally attempts to show how to set up containerized GenAI deployments for different runtimes (i.e., Podman, Kubernetes, Apptainer), different environments (i.e., HPC, cloud-based), and vendor-specific accelerators.

While these studies provide valuable insights into inference deployment, the existing literature exhibits several notable gaps. Most research focuses on single deployment environments or specific vendor ecosystems, lacking comprehensive comparisons across heterogeneous HPC and cloud infrastructures. Additionally, limited attention has been given to air-gapped deployment scenarios, cross-vendor accelerator compatibility, and the practical challenges of multi-runtime orchestration in production environments within HPC centers.

Our work explores and begins to address these limitations by presenting a converged computing environment architecture, integrating with existing HPC platforms, and covering the required steps for deploying containerized GenAI services across diverse computing platforms and accelerator types. Our experience suggests that that many of the container deployment differences between environments could be automated in order to reduce the burden on HPC center users.

## 6 Conclusion

This paper has presented our experiences deploying GenAI workloads within our HPC center. We have described the converged computing architecture we are pursuing, which combines HPC platforms, Kubernetes, site-wide object storage, and container registries to support GenAI applications. We have presented an end-to-end case study deploying the vLLM inference server using containers across multiple platforms, demonstrating both single and multi-node inference deployments and performance results. We have highlighted our observations and lessons learned, discussing opportunities for simplifying the deployment of containerized GenAI applications within HPC centers. Building on these insights, we have begun prototyping container deployment tools to address the identified challenges and help HPC users more easily deploy containers.

## Acknowledgments

Sandia National Laboratories is a multi-mission laboratory managed and operated by National Technology & Engineering Solutions of Sandia, LLC (NTESS), a wholly owned subsidiary of Honeywell International Inc., for the U.S. Department of Energy's National Nuclear Security Administration (DOE/NNSA) under contract DE-NA0003525. This written work is authored by an employee of NTESS. The employee, not NTESS, owns the right, title and interest in and to the written work and is responsible for its contents. Any subjective views or opinions that might be expressed in the written work do not necessarily represent the views of the U.S. Government. The publisher acknowledges that the U.S. Government retains a non-exclusive, paid-up, irrevocable, world-wide license to publish or reproduce the published form of this written work or allow others to do so, for U.S. Government purposes. The DOE will provide public access to results of federally sponsored research in accordance with the DOE Public Access Plan.

## A  Artifact Description (AD) Appendix

### A.1  Artifact Identification

Our work presents an end-to-end containerized workflow for deploying the vLLM inference server on HPC and Kubernetes platforms. Example scripts for deploying this workflow are provided in an accompanying Git repository at https://github.com/ktpedre/canopie25-paper-artifacts. This repository also includes the raw results presented in the paper and the Gnuplot scripts used to generate the plots.

### A.2  Reproducibility of Experiments

Our workflow is detailed in the text of the paper, and the example scripts provided in the Git repository enable readers to recreate the workflow on their own hardware platforms.

Reproducing the results will require approximately 10,000 minutes (167 hours) of compute node time. Compute nodes must provide at least 4x 80 GiB GPUs per node to run the Llama 4 Scout and Llama 3.1 405B models. The 4-bit quantized version of the Llama 4 Scout model can run on nodes with 2x 80 GiB GPUs per node.

Our experience indicates that run-to-run variability is very low. All models used are publicly available, and the vLLM inference server is open source. Additionally, all container images used are also public.

The expected results from the workflow, the inference performance results, are provided in data files located in the accompanying Git repository. Users can add their own results to these files and update the Gnuplot scripts to plot their results alongside those presented in the paper.